\documentclass{goeproc}

\usepackage{shortvrb}
\MakeShortVerb{\|}
\def\pdfLaTeX{pdf\kern.06em\LaTeX}
\begin{document}

\title{Sunspot models with bright rings}

\author[1,2*]{L.~L.~Kitchatinov}
\author[2]{G. R\"udiger}

\affil[1]{Institute for Solar-Terrestrial Physics, Irkutsk, Russia}
\affil[2]{Astrophysikalisches Institut Potsdam, Germany}
\affil[*]{\textit{Email:} lkitchatinov@aip.de}

\runningtitle{Sunspot models with bright rings}
\runningauthor{L.~L.~Kitchatinov and G.~R\"udiger}

\firstpage{1}

\maketitle

\begin{abstract}
A theoretical sunspot model is provided including
magnetic suppression of the diffusivities and also a strong
stratification of density and temperature. Heat diffusion alone
with given magnetic field and zero mean flow only produces  (after a very long relaxation time)  dark
spots without any bright ring.
Models with full dynamics of both
field and flow, however,   provide rings and also the observed
correlation of ring temperature excess and the spot size. The
rings are formed as the result of   heat
transport by the resulting flow system  {\em and}  increased thermal
diffusivity due to reduced magnetic quenching around spots.

 \end{abstract}
\section{Introduction}
Measurements have shown the
existence of bright rings around sunspots (Bonnet et al.~1978; Rast et al.~1999, 2001).
The rings appear one spot
radius beyond the spots and are reported to be about 10\,K warmer than
the  photosphere.  There is a  clear trend that
larger spots have brighter rings. We shall discuss in the following whether simple  mean-field models of more or less flat sunspots do develop such rings or not.

\section{A simple model}
The model concerns a horizontal layer with top and bottom at
depths $d_\mathrm{top}$ and $d_\mathrm{bot}$ below the
photosphere. In our computations $d_\mathrm{top} = 1$~Mm is
fixed, slightly beneath the typical depth of Wilson depression.
The fluid is assumed to be a perfect gas. Density and temperature
at the top ($\rho_\mathrm{top}, T_\mathrm{top}$) and bottom
($\rho_\mathrm{bot}, T_\mathrm{bot}$) boundaries are prescribed
with the solar structure model by Stix \& Skaley (1990).
The reference atmosphere is approximated by adiabatic profiles,
\begin{equation}
   T = T_\mathrm{top} + \frac{g}{C_\mathrm{p}}\left( D - z\right) ,\quad \quad \quad
   \rho = \rho_\mathrm{top}\left(\frac{T}{T_\mathrm{top}}
   \right)^\frac{1}{\gamma - 1} .
   \label{1}
\end{equation}
Here the gravity is $g = 2.74\cdot
10^4$~cm\,s$^{-2}$, $D = d_\mathrm{bot} - d_\mathrm{top}$ is the
layer depth (here 15 Mm), $z$ is the vertical coordinate, $C_\mathrm{p}$ and $\gamma$ are
specific heat and adiabaticity index defined by the condition
that the bottom temperature and density are exactly reproduced,
i.e.
\begin{equation}
   C_\mathrm{p} = \frac{gD}{T_\mathrm{bot}-T_\mathrm{top}} ,\quad \quad \quad \quad
   \gamma = 1 + \frac{\ln ( T_\mathrm{bot}/T_\mathrm{top})}
   {\ln (\rho_\mathrm{bot}/\rho_\mathrm{top})} .
   \label{2}
\end{equation}
The entropy equation with $S = C_\mathrm{v}\log \left( P/\rho^{\gamma}\right)$ is
\begin{equation}
   \rho T \left(\frac{\partial S}{\partial t}
   + {\vec u}\cdot{\vec\nabla} S\right) =
   {\vec\nabla}\cdot\left(\rho T \chi {\vec\nabla} S\right).
   \label{5}
\end{equation}
The sunspot darkness is usually explained in terms of convective
heat transport suppressed by magnetic field. Accordingly, the eddy
diffusivities in the model depend on the magnetic field (cf.
R\"udiger \& Kitchatinov 2000).
This dependence describes a steady decrease of the turbulent
diffusivities with the magnetic field  amplitude. Hence $
   \chi = \chi_\mathrm{T}\varphi\left(\beta\right),
$
where $\chi_\mathrm{T}$ is nonmagnetic diffusivity and the quenching function
\begin{equation}
   \varphi\left(\beta\right) =
   \frac{3}{8\beta^2}\left(\frac{\beta^2 -1}{\beta^2 + 1}
   + \frac{\beta^2 + 1}{\beta}\mathrm{arctan}\,\beta\right)
   \label{7}
\end{equation}
depends on the field strength normalized to the energy
equipartition value, $\beta = B/(\sqrt{\mu_0\rho}\, u')$, $u'$ is
the rms turbulent velocity.
Any
anisotropy of the eddy diffusivities is neglected.

The depth profiles of the equipartition field, $B_\mathrm{eq}
=\sqrt{\mu_0\rho}\, u'$, and the (nonmagnetic) diffusivities are
written in accordance with the  mixing-length approximation,
which yields
\begin{eqnarray}
   \chi_\mathrm{T}\left( z\right) = \chi_0 \left(\frac{T}
   {T_\mathrm{top}}\right)^{\frac{3\gamma - 4}{3\left(\gamma - 1\right)}}, \ \ \ \ \ \ \ \ \ \ \ \ \ \
   B^2_\mathrm{eq} = B^2_0\left(\frac{T}{T_\mathrm{top}}\right)^\frac{1}{3\left(\gamma - 1\right)},
  \label{9}
\end{eqnarray}
where $T$ has been defined in (\ref{1}) while  $\chi_0$ and $B_0$
are the thermal diffusivity and the equipartition field on the top
boundary. A marginal value for thermal convection,
$\chi_0 = 1.4 \cdot 10^{13}$\,cm$^2$/s, was taken for the
diffusivity, and $B_0=500$ Gauss. The density runs from $2.02\times
10^{-6}$ g/cm$^3$ at the top to $1.8\times 10^{-3}$ g/cm$^3$ at
the bottom, and the temperature varies from 14,000 K to 120,000 K.

The horizontal boundaries are  stress-free and impenetrable. For the magnetic field a vacuum
condition is used for the top while the field is assumed vertical
on the bottom. Thermal conditions are the black-body radiation on
the top and constant heat flux ($F_0 = 6.27\cdot
10^{10}$\,g\,s$^{-3}$) at the bottom. If wall boundaries are used
we assume zero stress, zero normal velocity, and superconductor
outside.

\begin{figure}[htb]
   \mbox{
  \includegraphics[height=3.12cm]{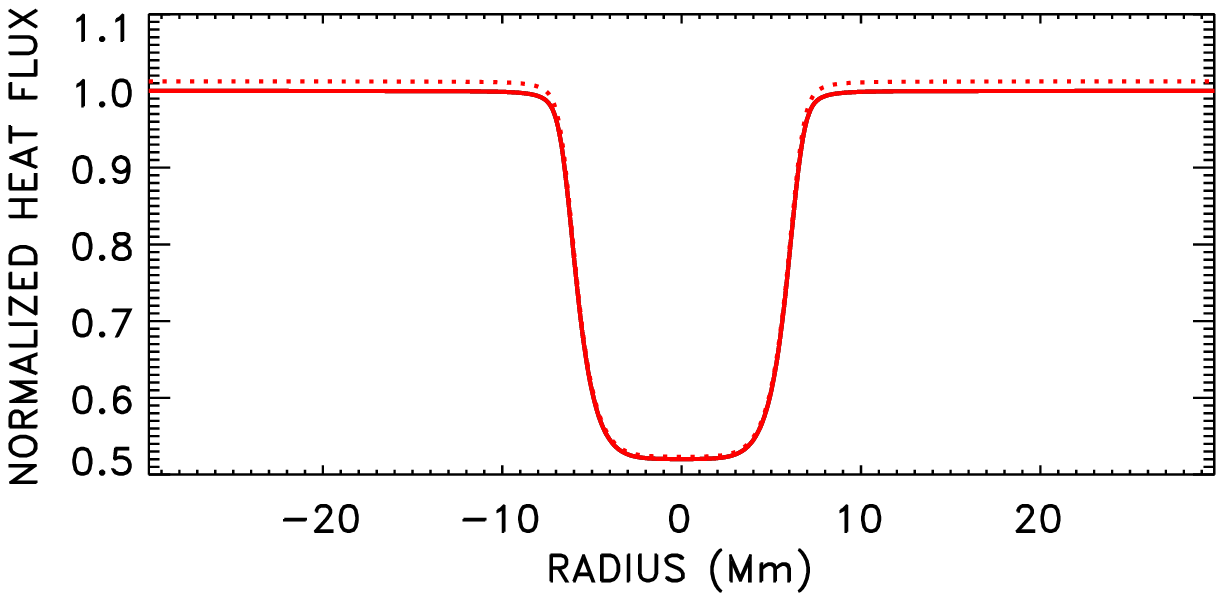}\hfill
   \includegraphics[height=3.12cm]{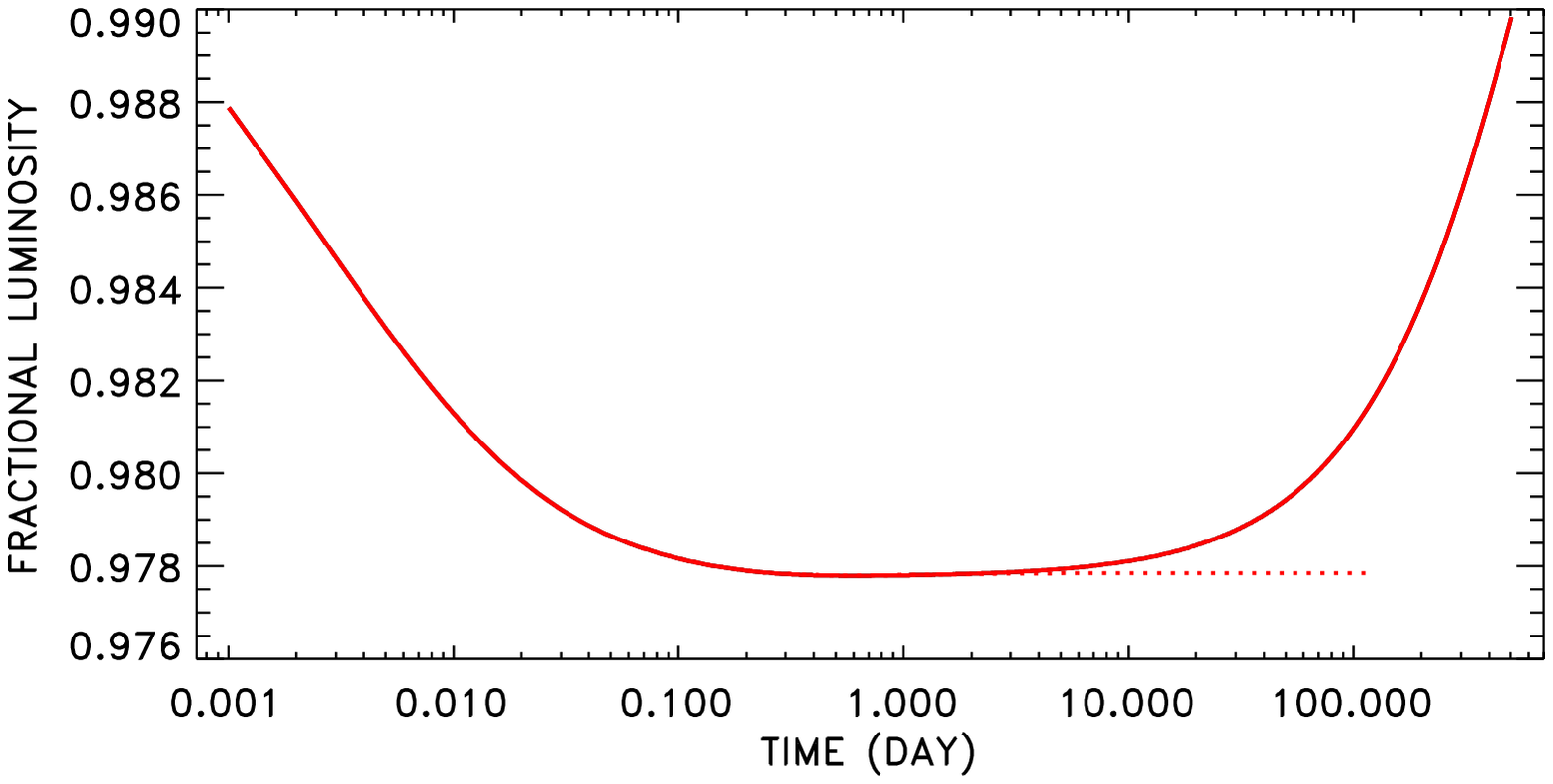}}
  \caption{Left: Profiles of surface brightness normalized to $F_0$
            after 1 day (solid) and after about
        500 days (dotted) for  vanishing heat flux
        across the wall boundary. Right: Time dependence of the normalized total irradiance at the top.
            The  thermal
        conditions on the side wall are $F_r=0$ (solid) and $\partial T/\partial t=0$ (dotted).  }
   \label{f3}
\end{figure}

A simplified model was used to probe the possibility of reproducing bright
rings by heat transport alone. In this model the velocity is put
to zero, the spot-like structure of the magnetic field is
prescribed and stationary  and the diffusion
equation for the entropy is solved.
It is a rather flat spot with an aspect ratio $a=2$. The spot of
this model has a magnetic flux of $3\cdot 10^{21}$\,Mx and a
depth of about 10\,Mm.

The surface brightness at the initial state is uniform and equals
$F_0$. The brightness of the region occupied by magnetic field
decreases  due to magnetic quenching of the thermal diffusivity.
The normalized surface brightness approaches the profile shown by
the solid line in Fig.\,\ref{f3} (left panel) already after
several hours and varies slowly after this time. The profile does
not show any bright ring.

The further evolution depends on the thermal conditions imposed
on the wall boundary. The boundary condition (zero heat flux
across the boundary) ensures that in the steady state the total
irradiance from the surface  equals the heat supply at the bottom
($\pi R^2 F_0$). In the steady solution the heat flux blocked by
the spot can reemerge {\em for flat spots} as a bright ring
(Eschrich \& Krause 1977).
 Figure\,\ref{f3} (right) shows that
the luminosity does indeed recover after the initial reduction
but it needs a very long time which is of the order of the Kelvin-Helmholtz time
($\tau_\mathrm{KH} \sim 10$~yrs for our model).

The reason for this long time scale  is the  high heat capacity of
the solar interior. The time passed before the rise in the surface
irradiance can be seen is much longer than lifetimes of sunspots.
Steady solutions of the heat transport equation are thus not
relevant for sunspots. 
Also flatter spot models do not show any bright rings.

\section{A new MHD sunspot model}\label{sec:demo}
For a more consistent MHD model for sunspots we work with  the anelasticity condition,   ${\vec\nabla}\cdot (\rho{\vec u})=0$, and  the complete momentum equation  together with the induction equation
\begin{equation}
   \frac{\partial{\vec B}}{\partial t} = {\vec\nabla}\times\big(
   {\vec u}\times{\vec B} - \eta{\vec\nabla}\times{\vec B}\big)\,.
   \label{4}
\end{equation}
\begin{figure}[htb]
   \center\includegraphics[height=5.5cm,width=7cm]{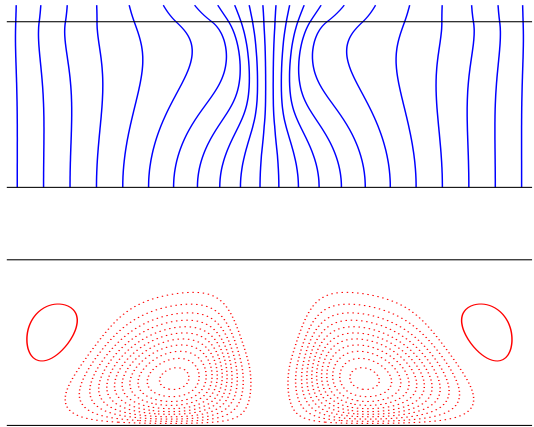}
   \caption{Field lines (top) and streamlines (bottom) after 3 days
            for an initial field of 600\,Gauss.
        The dotted streamlines mean the flow convergent
        on the top and divergent on the bottom. The depth of the
        simulated spot is comparable to its diameter.
        The surface flow is convergent near the spot but
        divergent at larger distances.
              }
   \label{f5}
\end{figure}

\noindent
Amplitudes
of the initial uniform field of several hundred Gauss are considered. Magnetic diffusivity and eddy viscosity are assumed to follow the same magnetic quenching expressions as  in Eq. (\ref{7}).

The vertical size of the simulated spots is comparable to their
diameter. The resulting mean `meridional circulation' is convergent on the top
near the spots and it makes downdrafts beneath the spots in agreement
with results of heliotomography (Fig.~\ref{f5}, bottom). The amplitude of the
flow is about 800~m\,s$^{-1}$. The flow direction becomes
divergent at larger radial distances.
The field strength in the spot is about 2700 Gauss. Brightness
and field strength are almost uniform in the central parts of the
spots but they change rapidly with radius beyond the \lq umbra'.

The radial heat-flux profiles show bright rings around the spots.  Our rings are somewhat brighter than the observed rings.

\begin{figure}[htb]
   \center\includegraphics[width=10cm,height=3.5cm]{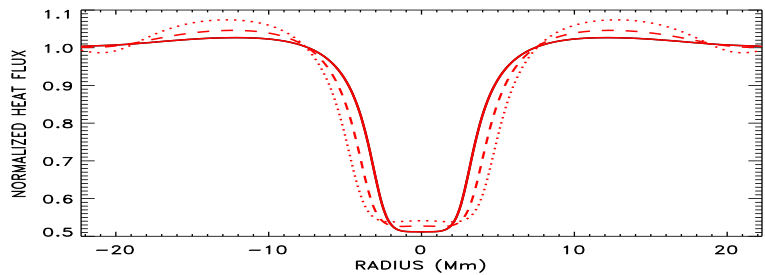}
   \caption{Surface brightness normalized to $F_0$ for the
            time of 3 days in the runs for initial fields
        of 400 (full line), 500 (dashed), and 600\,Gauss (dotted).
        All runs show bright rings which are more
        pronounced for larger spots.
              }
   \label{f7}
\end{figure}

A comparison of Figs.\,\ref{f5} and  Fig.\,\ref{f7}
shows that the position of the ring maximum  is at the
radius where the horizontal surface flow changes its direction
and an upwards directed  subsurface flow appears. At the same
radius the surface field strength has a minimum. Hence, the upward flow carries extra heat
from beneath {\em and} the reduced field strength increases the
diffusive heat flux due to reduced magnetic quenching of the
thermal diffusivity (\ref{7}).
If  the flow is  switched off  to probe its contribution to the bright rings they do not disappear.  We conclude that both the flow and the  reduced diffusivity quenching contribute to the resulting bright rings.
The contribution by the circulation should basically be steady. Observations can help to decide between the two possible explanations for the  rings by finding out whether the rings around big spots are decaying on a time scale of some days or not.

\end{document}